\documentclass[useAMS,usenatbib]{mn2e}
\usepackage{epsfig}

\newcommand{\mwfrac}[1]{{\langle x_{#1}}\rangle_{\rm M}}
\newcommand{\meanAV}{\langle A_{\rm V} \rangle}
\newcommand{\meanW}{\langle W_{\rm_{CO}} \rangle}
\newcommand{\meanHt}{\langle {\rm N_{H_{2}}} \rangle}
\def\simless{\mathbin{\lower 3pt\hbox
   {$\rlap{\raise 5pt\hbox{$\char'074$}}\mathchar"7218$}}}
\def\simgreat{\mathbin{\lower 3pt\hbox
   {$\rlap{\raise 5pt\hbox{$\char'076$}}\mathchar"7218$}}}

\title[H$_2$ Versus CO Abundance]{On the Relationship Between Molecular Hydrogen and Carbon
  Monoxide Abundances in Molecular Clouds}
\author[Glover \& {Mac Low}]{S.~C.~O.~Glover$^{1}$, M.-M.~{Mac Low}$^2$ \\
$^1$Zentrum f\"ur Astronomie der Universit\"at Heidelberg, Institut f\"ur Theoretische
Astrophysik, Albert-Ueberle-Str.\ 2, 69120 Heidelberg \\
$^2$Department of Astrophysics, American Museum of Natural History, Central Park
 West at 79th Street, New York, NY 10024 \\
 {\tt email:} sglover@ita.uni-heidelberg.de, mordecai@amnh.org
}

\begin{document}

\maketitle

\begin{abstract}
  The most usual tracer of molecular gas is line emission from CO.
  However, the reliability of that tracer has long been questioned in
  environments different from the Milky Way.  We study the
  relationship between H$_2$ and CO abundances using a fully dynamical
  model of magnetized turbulence coupled to a chemical network
  simplified to follow only the dominant pathways for H$_2$ and CO
  formation and destruction, and including photodissociation using a
  six-ray approximation. We find that the abundance of H$_2$ is
  primarily determined by the amount of time available for its
  formation, which is proportional to the product of the density and
  the metallicity, but insensitive to photodissociation.
  Photodissociation only becomes important at extinctions under a few
  tenths of a visual magnitude, in agreement with both observational
  and prior theoretical work.  On the other hand, CO forms quickly,
  within a dynamical time, but its abundance depends primarily on
  photodissociation, with only a weak secondary dependence on H$_2$
  abundance.  As a result, there is a sharp cutoff in CO abundance at
  mean visual extinctions $A_{\rm V} \simless 3$.  At lower values of
  $A_{\rm V}$ we find that the ratio of H$_2$ column density to CO
  emissivity $X_{\rm CO} \propto A_{\rm V}^{-3.5}$.  This explains the
  discrepancy observed in low metallicity systems between cloud masses
  derived from CO observations and other techniques such as infrared
  emission. Our work predicts that CO-bright clouds in low metallicity
  systems should be systematically larger or denser than Milky Way
  clouds, or both.  Our results further explain the narrow range of
  observed molecular cloud column densities as a threshold effect,
  without requiring the assumption of virial equilibrium.

\end{abstract}


\section{Introduction}
Observed star formation takes place within giant molecular clouds  
(GMCs), so understanding how these clouds form and evolve is a key step towards
understanding star formation on galactic scales. The main chemical constituent of
any GMC is molecular hydrogen (H$_{2}$). However, it is extremely difficult to directly 
observe this molecular hydrogen {\em in situ}.  
Radiative transitions in H$_{2}$ are weak, owing to the 
molecule's lack of a permanent dipole moment. Moreover, the lowest lying 
rotational energy levels of H$_{2}$ are widely spaced, and so are very rarely 
excited in gas with the temperatures typical of GMCs, $T \sim 10$--20~K. For
this reason it is common to use emission from carbon monoxide (CO), the
second most abundant molecule in GMCs, as a proxy for the H$_{2}$.

In order to use CO as a proxy for H$_{2}$, however, it is necessary to
understand the relationship between the distributions of these two
molecules. Both are readily dissociated by the absorption of
ultraviolet photons with energies below the Lyman limit of atomic
hydrogen, so in low density, low
extinction gas such as the warm neutral component of the interstellar
medium (ISM),
the abundances of both molecules will
be small. However, the two molecules form rather
differently: H$_{2}$ forms predominantly on the surface of dust
grains \citep{gs63}, while CO forms almost exclusively in the gas
phase, via any one of a number of chains of ion-neutral or
neutral-neutral reactions (see e.g.\ \citealt{sd95} for a useful
summary of CO formation chemistry).  Moreover, H$_{2}$ can protect
itself from ultraviolet radiation via self-shielding, which becomes
effective for relatively low H$_{2}$ column densities
\citep{db96}. The corresponding process for CO is less effective
\citep{lee96}, requiring a higher column density, and the low
abundance of carbon relative to hydrogen in the ISM means that the
required column density typically corresponds to a situation in which
a large fraction of the available carbon is already locked up in the
form of CO. For these reasons, we would expect the CO/H$_{2}$
ratio to vary through a cloud, and this expectation is
confirmed by detailed models of slab-like or spherical clouds
\citep[e.g.][]{pdr07}.  These find that the transition from atomic
hydrogen to molecular hydrogen occurs at a point closer to the cloud
surface than the transition from ionized carbon, via neutral atomic
carbon, to CO.

Although spatial variations in the CO/H$_{2}$ ratio within a GMC
would appear to make it difficult
to use CO as a proxy for H$_{2}$, observations of Galactic GMCs show that in fact there
appears to be a good correlation between the integrated intensity of the $J = 1
\rightarrow 0$ rotational transition line of $^{12}$CO and the H$_{2}$ column density
\citep[see e.g.][]{dick78,sand84,sol87,sm96,dame01}. A number of independent studies 
have shown that GMCs in the Galactic disk all have CO-to-H$_{2}$ conversion factors 
that are approximately 
\begin{equation}
X_{\rm CO} = \frac{N_{\rm H_{2}}}{W_{\rm CO}} \simeq 2 \times 10^{20} {\rm cm^{-2} \: K^{-1} \: km^{-1} \: s},
\end{equation}
where $W_{\rm CO}$ is the velocity-integrated intensity of the CO $J = 1 \rightarrow 0$ 
emission line, averaged over the projected area of the GMC, and $N_{\rm H_{2}}$ is the 
mean H$_{2}$ column density of the GMC, averaged over the same area.
Although the former can be directly observed, 
the latter cannot. However, for nearby clouds, it can be inferred from measurements of the
diffuse $\gamma$-ray flux produced by interactions between high energy
cosmic rays and atomic hydrogen, atomic helium and H$_{2}$. The $\gamma$-ray
flux along a given line of sight depends on the total hydrogen column density along that line 
of sight. Since the atomic hydrogen column density can be measured via its 21~cm emission,
the H$_{2}$ column density can be inferred. As the values for $X_{\rm CO}$ obtained
in this way are consistent with those obtained by assuming that the GMCs are in virial 
equilibrium and using the observed linewidth-size relation to compute the cloud mass
\citep[e.g.][]{sol87}, it is generally accepted that CO emission is indeed a good proxy for H$_{2}$
mass in nearby GMCs.

However, the issue of the environmental dependence of $X_{\rm CO}$ remains highly
contentious. In gas with a lower metallicity, or a higher ambient UV radiation field,
CO photodissociation will be more effective, and the amount of CO in the cloud will be
smaller, with the CO-rich gas occupying a smaller volume than in the 
Galactic case (e.g.\ \citealt{mb88}, Molina et~al., in prep.). The mean H$_{2}$ abundance 
may also be smaller, but the greater role played by H$_{2}$ self-shielding means that
we do not expect the H$_{2}$ abundance to be nearly as sensitive to changes in the metallicity.
Because of this, there are good theoretical reasons to expect the relationship between
CO emission and H$_{2}$ mass to change as we change the metallicity. However,
observational efforts to test this yield inconsistent results.

Measurements of $X_{\rm CO}$ that assume that extragalactic GMCs are in virial equilibrium
and use a virial analysis to determine the cloud mass generally find values for $X_{\rm CO}$ 
that are similar to those obtained in the Milky Way, with at most a weak metallicity dependence
\citep{wilson95,ros03,bolatto08}. 
On the other hand, measurements that constrain GMC masses using other
techniques that do not depend on the CO emission, such as by measuring the far-infrared
dust emission, consistently find values for  $X_{\rm CO}$ that are much larger than the Galactic 
value and that are suggestive of a strong metallicity dependence 
\citep{israel97, rubio04, leroy07, leroy09}.

Numerical simulations provide us with one way to address this observational dichotomy.
If we can understand the distribution of CO and H$_{2}$ in realistic models of
GMCs, then we may begin to understand why the different types of observation give such
different results. However, until very recently, the ability of simulations to address this issue
has been quite limited. Sophisticated treatments of gas and grain chemistry, radiative heating
and cooling, and radiative transfer have been developed to model GMCs \citep[see, for instance, 
the paper by][which compares results from a number of popular codes]{pdr07}, but the very
complexity of these models limits their applicability to highly simplified geometries: typically
the adopted cloud models are one-dimensional, assuming either spherical symmetry 
\citep[e.g.][]{kosma96} or a semi-infinite, uniform slab \citep[e.g.][]{meudon06}.

Moreover, the modelling often assumes that the
chemistry of the clouds is in equilibrium, which is valid only if all of the chemical timescales
are much shorter than any dynamical timescale associated with the GMCs. Since real GMCs
are observed to be highly inhomogeneous, and to be dominated by supersonic turbulent 
motions \citep[see e.g.][and references therein]{mk04}, the applicability of the results from these 
simple spherical or slab models to real clouds is open to question. On the other hand, previous 
attempts to accurately model the turbulent dynamics of GMCs, and the inhomogeneous and 
intermittent density structure that is created by this turbulence typically have avoided modelling 
the cloud chemistry, rendering them also of limited use in addressing this question. The few 
studies that have attempted to model both the turbulence and the chemistry self-consistently 
\citep[e.g.][]{jou98,les07,god09} have typically done so by reducing the dimensionality of the
problem. For instance, \citet{god09} present results on the chemical evolution of gas passing
through magnetized, turbulent vortices, using two-dimensional simulations of these vortices,
and then construct models for lines of sight through diffuse clouds by summing up the contributions
from a number of different vortices. However, it is unclear whether the results obtained in this 
fashion are the same as those that would be obtained using a fully three-dimensional model
for the turbulence.

In a previous paper \citep[][hereafter Paper I]{glo10}, we presented the first results from a project 
that aims to combine detailed chemical modelling with three-dimensional magnetohydrodynamical
turbulence simulations in order to self-consistently model both the chemistry  and the 
turbulent dynamics of the gas within a GMC. We showed that it is now computationally feasible
to attempt this. Provided one makes a few simplifying assumptions 
regarding the extent of the chemistry to be treated, and the treatment of the UV radiation field,
it is possible to model both the H$_{2}$ and the CO chemistry of a GMC with acceptable
accuracy within a moderate resolution dynamical simulation. Paper~I presented results
from a few trial simulations performed using only one set of cloud properties (mean density, 
metallicity, etc). In this paper, we present results from a much larger set of simulations that
examine the sensitivity of H$_{2}$ and CO formation to changes in the mean densities and
metallicities of GMCs, and explore the consequences of this for the CO-to-H$_{2}$ conversion
factor and the dynamical structure of the observed clouds.

In section~\ref{sims}, we discuss our numerical approach and the
initial conditions used for our simulations. In section~\ref{res}, we
present the main results of our simulations, and use them to derive
the approximate dependence of the CO-to-H$_{2}$ conversion factor on
cloud properties.  In section~\ref{dis}, we discuss the major
consequences of our findings. In section~\ref{cave}, we discuss a
few potential caveats regarding our approach, and show why they 
are unlikely to significantly affect our main results. Finally, in
section~\ref{summ}, we close with a brief summary.

\section{Simulations}
\label{sims}
\subsection{Numerical method}
Our simulations were performed using a modified version of the ZEUS-MP
magnetohydrodynamical code \citep{norman00,hayes06} fully described in
Paper~I.  Our modifications include the addition of a simplified
treatment of hydrogen, carbon and oxygen chemistry, composed of 218
reactions between 32 chemical species, together with a detailed atomic
and molecular cooling function. This treatment includes the formation 
of H$_{2}$ on the surface of dust grains, following a prescription taken 
from \citet{hm79}. The formation rate of H$_{2}$ per unit volume due to
grain surface reactions is written as
\begin{equation}
R_{\rm H_{2}} = \gamma_{\rm H_{2}} \left(\frac{{\rm Z}}{{\rm Z_{\odot}}} \right) 
n n_{\rm H} \: {\rm cm^{-3}} \: {\rm s^{-1}},
\end{equation}
where $n$ is the number density of hydrogen nuclei, $n_{\rm H}$ is the
number density of atomic hydrogen, ${\rm Z}$ is the metallicity of the gas,
and $\gamma_{\rm H_{2}}$ is the rate coefficient for H$_{2}$ formation,
given by
\begin{equation}
\gamma_{\rm H_{2}} = \frac{3.0 \times 10^{-17} T_{2}^{1/2} f_{\rm a}}{1.0 + 
0.4 (T_{2} + T_{\rm d,2})^{1/2} + 0.2 T_{2} + 0.08 T_{2}^{2}},
\end{equation}
where $T_{2} = T / 100 \: {\rm K}$ is the gas temperature in units of 100~K,
$T_{\rm d,2} = T_{\rm d} / 100 \: {\rm K}$ is the dust temperature in units of 
100~K, and where
\begin{equation}
f_{\rm a} = \frac{1}{1 + 10^{4} \exp \left(-600 / T_{\rm d} \right)}.
\end{equation}
For simplicity, we fix the dust temperature at 10~K in all of the simulations
presented in this paper. Aside from this reaction, we include no other grain 
surface chemistry in our current models.

We model the photodissociation of H$_{2}$ using prescriptions for dust
shielding and H$_{2}$ self-shielding taken from \citet{db96}. For CO 
photodissociation, we account for dust shielding, CO self-shielding
and shielding by H$_{2}$, using shielding functions taken from \citet{lee96} for
the last two terms. To compute the column densities of dust, H$_{2}$ and CO
required by this treatment of photodissociation, we use the six-ray approximation
introduced in \citet{gm07a}. In this approximation, we compute 
photochemical rates in each zone in our simulation volume by averaging over 
the six lines of sight that lie parallel to the coordinate axes.

\subsection{Initial conditions}
All of our simulations begin with an initially uniform gas distribution,
with mean hydrogen nucleus number density $n_{0}$. The initial temperature
of the gas is 60~K. However, because the cooling time of the gas is significantly 
shorter than the dynamical time in all but our lowest density runs,  the gas temperature 
quickly adjusts itself until the gas is close to thermal equilibrium, and hence our results 
are largely insensitive to our choice for the initial temperature. The initial velocity field 
is turbulent, with uniform power between wavenumbers $k = 1$ and $k = 2$, and with 
an initial rms velocity of 5~${\rm km}
\: {\rm s^{-1}}$. We drive the turbulence so as to maintain approximately the 
same rms velocity throughout the simulation, following the method
described in \citet{mkbs98} and \citet{ml99}.

We assume that the gas is magnetized, with an initially uniform magnetic field 
strength $B_{0} = 5.85 \mbox{ }\mu {\rm G}$ and a field that is initially oriented parallel to 
the $z$-axis of the simulation. In \citet{gm07b} we showed that the timescale
for H$_{2}$ formation in a turbulent, magnetized cloud is relatively insensitive 
to the value chosen for the magnetic field strength, and hence we do not expect
our results to be particularly sensitive to our choice of $B_{0}$. The particular 
value that we have chosen is the same as we adopted in Paper~I, and is
motivated by the 21~cm observations by \citet{ht05} that find a median magnetic
field strength of $\sim 6 \mu {\rm G}$ for the cold neutral medium.

We assume that the gas has a uniform metallicity ${\rm Z}$, and that the 
initial fractional
abundances (by number) of carbon and oxygen relative to hydrogen are
given by
$x_{\rm C^{+}}  = x_{\rm C, tot} = 1.41 \times 10^{-4} ({\rm Z} / {\rm Z_{\odot}})$ and 
$x_{\rm O}   =x_{\rm O, tot} = 3.16 \times 10^{-4} ({\rm Z} / {\rm Z_{\odot}})$, where $x_{\rm C, tot}$
and $x_{\rm O, tot}$ refer to carbon and oxygen in all forms (ionized, neutral, or incorporated
into molecules). For solar metallicity, this corresponds to the values
given in \citet{sem00}. 
As in \citet{gm07b}, we ignore the effects of metals other than C or O, in order to minimize the 
computational requirements of our simulations. We assume
a standard dust-to-gas ratio for our solar metallicity runs, and also assume that
in our lower metallicity runs, the dust-to-gas ratio scales linearly with metallicity.
We assume that the dust has an extinction curve characterized by $R_{\rm V}
= 3.1$, and that this is independent of metallicity. We also fix the dust temperature
at $10 \: {\rm K}$ in every run. 


We adopt a fixed cosmic ray ionization rate $\zeta = 10^{-17} \: {\rm s^{-1}}$. 
For the majority of our simulations, we use the same spectral shape and
normalization for the interstellar radiation field as in \citet{dr78}. For brevity,
when we refer to the strength of the radiation field in a simulation, we typically
give it in terms of $G_{0}$, the ratio of the field strength relative to the strength
of the Draine field. Thus, in the majority of our simulations, $G_{0} = 1.0$. We
also performed a few simulations without a radiation background, i.e.\ with
$G_{0} = 0.0$. The effects of increasing $G_{0}$ to much larger values are
examined in a separate paper (Glover et~al., in prep).

We perform our simulations in a cubical volume of side $L$, to which we apply 
periodic boundary conditions. As we have previously noted in Paper~I, the
use of periodic boundary conditions for the gas together with what are effectively
non-periodic boundary conditions for the radiation is not self-consistent, and we
adopt this arrangement purely for the computational convenience that it offers.
We do not include the effects of self-gravity in our current simulations, but intend
to explore them in future work.

We have performed simulations with a range of different initial densities $n_{0}$
and  metallicities ${\rm Z}$, as detailed in Table~\ref{tab:sims}. For most of these
simulations, we set $L = 20 \: {\rm pc}$, but we have also performed a few 
simulations with $L = 5 \: {\rm pc}$.  For all of these simulations, we adopt a
numerical resolution of $128^{3}$ zones, which we showed in Paper~I was
sufficient to yield converged results for the mass-weighted mean abundances of
CO and H$_{2}$, at least to the level of accuracy required here.

We evolved all of our runs until $t_{\rm end} = 1.8 \times 10^{14} \: {\rm s}
\simeq 5.7 \: {\rm Myr}$, corresponding to roughly three eddy turnover times for
the turbulence. In Paper~I, we showed that this gave sufficient time
for the density distribution, temperature distribution and the chemical abundance
of CO  all to reach a statistical steady state. We expect the same to be true 
for the majority of runs examined here. However, as we will see later, there is 
evidence that the CO abundance in some of our lowest density runs has not reached a 
steady state by $t = t_{\rm end}$. Therefore, in four of our lowest density runs
(runs n30, n100, n30-Z01 and n100-Z01), we did not stop at this point, but continued to 
evolve them until a time $t_{\rm ext} = 7.2 \times 10^{14} \: {\rm s} \simeq 22.8 \: {\rm Myr}$. 
We note,  however, that it is questionable whether real molecular clouds could live for so
long without undergoing localized gravitational collapse and forming stars, physics
that is not included in our current models.

\begin{table}
\caption{List of simulations \label{tab:sims}}
\begin{tabular}{lcccc}
\hline
ID & $n_{0}$ (cm$^{-3}$) & ${\rm Z / Z_{\odot}}$ & Box size (pc) & $G_{0}$ \\
\hline
n30 & 30 & 1.0 & 20 & 1.0 \\
n100 & 100 & 1.0 & 20 & 1.0 \\
n180 & 180 & 1.0 & 20 & 1.0 \\
n300 & 300 & 1.0 & 20 & 1.0 \\
n1000 & 1000 & 1.0 & 20 & 1.0 \\
n300-Z06 & 300 & 0.6 & 20 & 1.0 \\
n100-Z03 & 100 & 0.3 & 20 & 1.0 \\
n300-Z03 & 300 & 0.3 & 20 & 1.0 \\
n30-Z01 &30 & 0.1 & 20 & 1.0 \\
n100-Z01 &100 & 0.1 & 20 & 1.0 \\
n300-Z01 &300 & 0.1 & 20 & 1.0 \\
n1000-Z01& 1000 & 0.1 & 20 & 1.0 \\
n1000-Z003 & 1000 & 0.03 & 20 & 1.0 \\
n30-L5 & 30 & 1.0 & 5 & 1.0 \\
n100-L5 & 100 & 1.0 & 5 & 1.0 \\
n300-L5 & 300 & 1.0 & 5 & 1.0 \\
n1000-L5 & 1000 & 1.0 & 5 & 1.0 \\
n30-UV0 & 30 & 1.0 & 20 & 0.0 \\
n100-UV0 & 100 & 1.0 & 20 & 0.0 \\
n300-UV0 & 300 & 1.0 & 20 & 0.0 \\
n1000-UV0 & 1000 & 1.0 & 20 & 0.0 \\
\hline
\end{tabular}
\medskip
\\
Note: $G_{0}$ is the ultraviolet field strength in 
units of the \citet{dr78} field
\end{table}

\section{Results}
\label{res}

\subsection{Density and temperature structure of the gas}
\begin{figure}
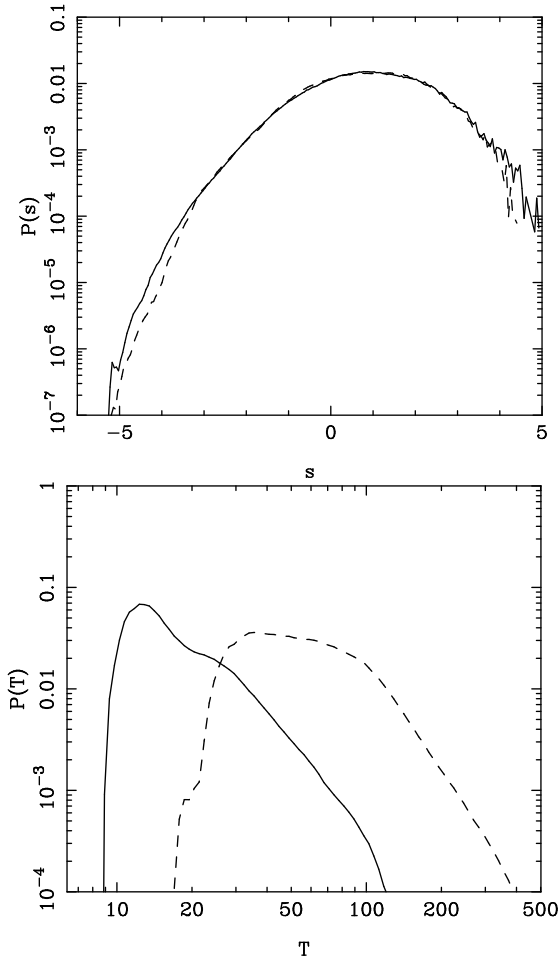

\centering
\epsfig{figure=f1a.eps,width=15pc,angle=270,clip=}
\epsfig{figure=f1b.eps,width=15pc,angle=270,clip=}
\caption{(a) Mass-weighted PDF of the logarithmic density contrast
$s = \ln (\rho / \rho_{0})$, where $\rho_{0}$ is the mean density of the gas,
plotted for runs n300 (solid line) and n30 (dashed line). Both of the PDFs display 
the log-normal shape characteristic of quasi-isothermal supersonic turbulence. 
(b) Mass-weighted temperature PDF for runs n300 (solid line) and 
n30 (dashed line). The characteristic temperature of the gas in run n30 is
much higher than in run n300, owing to a combination of the effects of the
lower gas density and the low CO abundance (see \S\ref{sec:abd} below), 
which render cooling much less effective, and the lower level of dust extinction, 
which significantly increases the importance of photoelectric heating. Note that
while the relatively sharp cutoff in the PDF for run n30 at $T < 30 \: {\rm K}$ 
appears to be real, the corresponding feature at $T < 10 \: {\rm K}$ in the PDF
for run n300 is not; the latter is a consequence of the fact that for technical
reasons,  our cooling  function cuts off at $T = 10 \: {\rm K}$ (see Paper I).
\label{nTpdf}}
\end{figure}

Before considering the molecular abundances produced in the various simulations, 
it is perhaps useful to briefly discuss the density and temperature distributions 
generated by the turbulence. As an illustrative example, we plot in Figure~\ref{nTpdf}a
the mass-weighted density probability distribution function (PDF) for runs n300 
and n30. The PDFs are plotted in terms of the dimensionless logarithmic density contrast 
$s \equiv \ln (\rho / \rho_{0})$, where $\rho_{0}$ is the mean density of the gas. It
is clear from the plot that the two PDFs have a similar form when plotted in dimensionless 
units. Both PDFs are roughly log-normal, with similar widths and means, as expected 
from previous studies of supersonic turbulence in interstellar gas \citep{pnj97}. 
Deviations from the log-normal form are apparent in the wings of the PDFs, where one would 
expect the effects of turbulent intermittency to be most pronounced \citep{krit07,schm09,fed10}.

It is also interesting to compare the temperature PDFs of the same two runs
(Figure 1b). If we do so, we find that the gas is systematically warmer in run
n30 than in run n300, owing to the increased effectiveness of photoelectric
heating and reduced effectiveness of CO cooling in the lower density run.
We find similar behaviour in our other runs, and so for reasons of brevity,
we will not plot their density and temperature PDFs here. In each case, we 
find a roughly log-normal density PDF, as expected, with a width that is correlated 
with the mean temperature of the gas. Reducing the density and/or the metallicity of 
the gas tends to increase its mean temperature, both by reducing the efficiency 
of cooling and by increasing the influence of photoelectric heating, and hence 
the lower density and lower metallicity runs have narrower density PDFs than 
the higher density or higher metallicity runs. This trend is also apparent if one 
looks at the clumping factor of the gas, defined as 
\begin{equation}
C = \frac{\langle\rho^{2}\rangle}{\langle\rho \rangle^{2}},
\end{equation}
which ranges from $C \simeq 4.5$ in run n1000-Z003 to $C \simeq 6.8$ in
run n1000, with the values from the other runs spread fairly evenly between
these two extremes. These values are consistent with those inferred from
models of some Galactic photodissociation regions \citep[e.g.][whose results
imply a mean value for $C$ of roughly 5 in W3 Main]{kramer04}, although
other regions show evidence for much higher values of $C$ \citep[e.g.][]{fp96}.

Although these results are not particularly surprising given our findings in
\citet{gm07b} and in paper I, they do serve to demonstrate that the density
structure produced by the turbulence is rather different from the structure 
that is often assumed in models of clumpy photodissociation regions (PDR). 
These models commonly assume that the density structure of a PDR can be
represented by a population of clumps (usually assumed to be spherical, but 
with densities and sizes that can vary from clump to clump), embedded in some 
much lower density interclump medium \citep[see e.g.][]{bht90,mt93,svd97,mook06}. 
However, the density 
distributions generated by the turbulence in our simulations do not fit comfortably 
into this picture. The fact that we find density PDFs that are smooth, single-peaked 
functions without even a hint of bimodal behaviour suggests that any attempt to 
partition the gas into high density clumps embedded in a much lower density medium 
will inevitably wind up being somewhat arbitrary, since there is no particular 
density or range of densities that one can pick out of the PDF as corresponding 
to a  `clump', and hence no obvious way to determine where a `clump' ends and
the interclump gas begins. This unfortunately makes it difficult to compare our 
results with those coming from these clumpy PDR models, but does also 
serve to emphasize the great importance of modelling the dynamics and 
chemistry of the gas in a  self-consistent fashion.

\subsection{Molecular abundances: dependence on extinction and gas density}
\label{sec:abd}

\begin{table}
\caption{Mass-weighted mean abundances of H$_{2}$ and CO 
at our default stopping time, $t = t_{\rm end}$ \label{tab:fractions}}
\begin{tabular}{lll}
\hline
ID & $\mwfrac{{\rm H_{2}}}$ &  $\mwfrac{{\rm CO}}$ \\
\hline
n30 & 0.351 & $9.31 (-9)$ \\
n100 & 0.689 & $2.97 (-6)$ \\
n180 & 0.847 & $1.92 (-5)$ \\
n300 & 0.943 & $6.41 (-5)$ \\
n1000 & 0.998 & $1.27 (-4)$ \\
n300-Z06 & 0.874 & $2.10 (-5)$ \\
n100-Z03 & 0.357 & $2.19 (-8)$ \\
n300-Z03 & 0.712 & $2.74 (-6)$ \\
n30-Z01 & 0.027 & $2.27 (-12)$ \\
n100-Z01 & 0.136 & $8.27 (-11)$ \\
n300-Z01 & 0.385 & $1.66 (-7)$ \\
n1000-Z01& 0.816 & $2.32 (-6)$ \\
n1000-Z003 & 0.449 & $6.08 (-8)$ \\
n30-L5 & 0.342 & $3.80 (-10)$ \\
n100-L5 & 0.817 & $1.26 (-8)$ \\
n300-L5 & 0.982 & $9.00 (-6)$ \\
n1000-L5 & 0.998 & $8.17 (-5)$ \\
n30-UV0 & 0.345 & $1.84 (-5)$ \\
n100-UV0 & 0.687 & $8.39 (-5)$ \\
n300-UV0 & 0.944 & $1.39 (-4)$ \\
n1000-UV0 & 0.998 & $1.41 (-4)$ \\
\hline
\end{tabular}
\end{table}

\begin{table}
\caption{Mass-weighted mean abundances of H$_{2}$ and CO
at the end of our extended runs \label{tab:fractions_ext}}
\begin{tabular}{lcc}
\hline
ID & $\mwfrac{{\rm H_{2}}}$ at $t_{\rm ext}$ &  $\mwfrac{{\rm CO}}$ at $t_{\rm ext}$ \\
\hline
n30 &  0.774 & $ 2.63 (-8)$ \\
n100 & 0.972  &  $1.55 (-5)$ \\
n30-Z01 & 0.055 & $ 6.78 (-12)$ \\
n100-Z01 & 0.445  & $ 2.84 (-10)$ \\
\hline
\end{tabular}
\end{table}

\begin{figure}
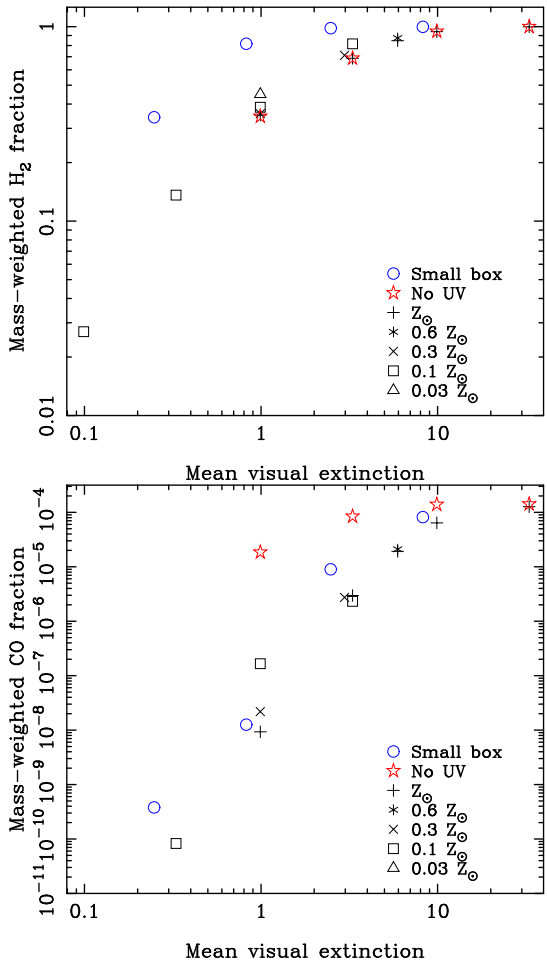

\centering
\epsfig{figure=f2a.eps,width=15pc,angle=270,clip=}
\epsfig{figure=f2b.eps,width=15pc,angle=270,clip=}
\caption{(a) Mass-weighted mean H$_{2}$ fraction, plotted as 
a function of the mean visual extinction $\meanAV$. Different
symbols are used for runs with different metallicities, box sizes
or UV field strengths, as outlined in the legend. The runs 
performed using an $L = 5 \: {\rm pc}$ box (blue circles) 
produce as much H$_{2}$ as the corresponding  $L = 20 \: {\rm pc}$ 
runs (crosses), and so the former are simply displaced to the
left of the latter by a factor of four. This fact, together with the fact
that the runs performed without a UV background (red stars)
follow the same apparent correlation as the runs with a UV 
background, provide convincing evidence that the apparent
correlation between $\mwfrac{\rm H_{2}}$ and $\meanAV$ is artificial: the
true correlation is between $\mwfrac{\rm H_{2}}$  and $n_{0} {\rm Z}$, and 
the apparent correlation with $\meanAV$ arises because
$\meanAV \propto n_{0} {\rm Z} L$.
(b) As (a), but for the mean mass-weighted CO fraction.
In this case, the $L = 5 \: {\rm pc}$ runs show the same 
correlation between the mean CO fraction and the 
mean extinction as the $L = 20 \: {\rm pc}$ runs, but the runs
performed without a UV background do not. This 
         suggests that 
the correlation between $\mwfrac{\rm CO}$ and $\meanAV$
         is 
real.
\label{mol-AV}
}
\end{figure}

\begin{figure}
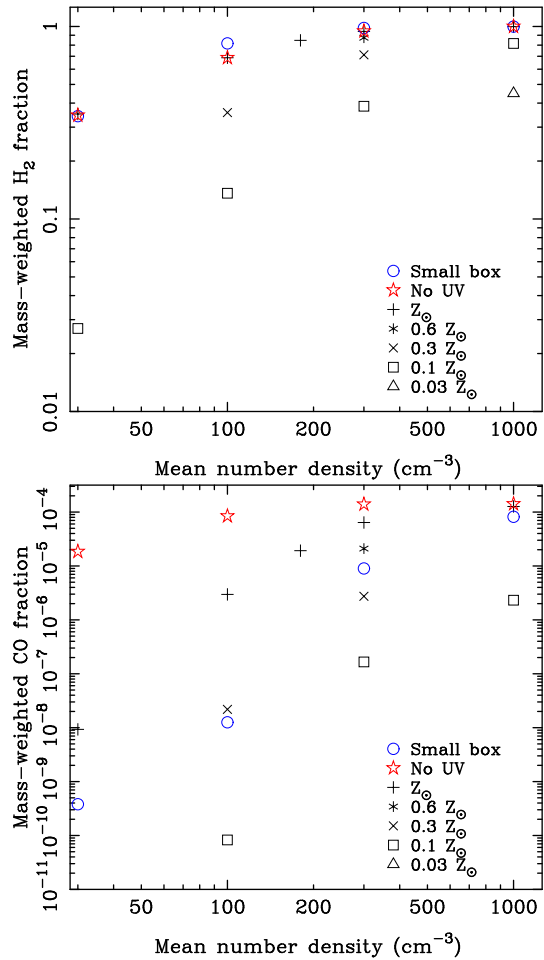

\centering
\epsfig{figure=f3a.eps,width=15pc,angle=270,clip=}
\epsfig{figure=f3b.eps,width=15pc,angle=270,clip=}
\caption{(a) Mass-weighted mean H$_{2}$ fraction, plotted as a function of
mean density $n_{0}$. We see a clear dependence on density at fixed 
metallicity, and a clear dependence on metallicity at fixed density, consistent
with 
$\mwfrac{\rm H_{2}}$ 
         correlating with
the product of density and metallicity.
(b) As (a), but for the mean mass-weighted CO fraction. Again, 
the CO fraction 
         depends on density 
at fixed metallicity, and
         depends on metallicity
at fixed density, 
          but now the CO fraction also depends strongly
on the box size $L$. 
          Density
by itself, without knowledge of $L$ or ${\rm Z}$, 
         predicts the CO fraction poorly.
\label{mol-n}
}
\end{figure}

Let us now turn our attention from the density and temperature of the gas to
its chemical content.
In Table~\ref{tab:fractions}, we list the mass-weighted mean abundances of
H$_{2}$ and CO, denoted $\mwfrac{{\rm H_{2}}}$  and $\mwfrac{{\rm CO}}$
respectively, at our default end-point $t = t_{\rm end} = 5.7 \: {\rm Myr}$, for each of the 
simulations that we have run. These mean abundances are defined as
\begin{equation}
\mwfrac{{\rm H_{2}}} =  \frac{2}{M} \Delta V \sum_{i, j, k} \frac{n_{\rm H_{2}}(i,j,k)}{n(i,j,k)} \rho(i,j,k),
\end{equation}
and
\begin{equation}
\mwfrac{{\rm CO}} = \frac{1}{M} \Delta V \sum_{i, j, k} \frac{n_{\rm CO}(i,j,k)}{n(i,j,k)} \rho(i,j,k),
\end{equation}
respectively, where $M$ is the total mass in the simulation,  $\Delta V$ is the volume of
a single grid zone, $n(i,j,k)$ is the number density  of hydrogen nuclei in zone $(i,j,k)$,  
$n_{\rm H_{2}}(i,j,k)$ and  $n_{\rm CO}(i,j,k)$ are the H$_{2}$ and CO number densities 
in the same zone, $\rho(i,j,k)$ is the mass density in that zone, and where we sum over
all grid zones. Note that the  inclusion of the factor of two in our definition of 
$\mwfrac{{\rm H_{2}}}$ is to ensure that when the hydrogen is fully molecular,
we will have $\mwfrac{{\rm H_{2}}} = 1.0$, rather than 0.5, as this will hopefully
be less confusing for the reader.

In Figure~\ref{mol-AV}a, we plot the values
of $\mwfrac{{\rm H_{2}}}$ at $t = t_{\rm end}$ as a function of the mean visual extinction, 
$\meanAV$, and in Figure~\ref{mol-AV}b, we give a similar plot of the values of 
$\mwfrac{{\rm CO}}$ at $t = t_{\rm end}$. The mean visual extinction 
$\meanAV$ is defined as
\begin{equation}
\meanAV = \frac{1}{N^{2}} \sum_{j,k} A_{{\rm V}, j, k} 
\end{equation}
where $N$ is the number of grid zones along one edge of the box,
$A_{{\rm V}, j, k}$ is the visual extinction along the line of sight running
parallel to the $x$-axis and with $y$ and $z$ coordinates $(j,k)$, and
where we sum over all sets of $(j,k)$ corresponding to zone centres.
Note that our choice to project along the $x$-axis is arbitrary, and that
projecting along the $y$ or $z$ axes would yield the same value of
$\meanAV$, which indeed is simply given by
\begin{equation}
\meanAV = f_{\rm conv} n_{0} L,
\end{equation}
where $f_{\rm conv}$ is the conversion factor between the column 
density of H nuclei and the visual extinction. This is given by
$f_{\rm conv} = 5.348 \times 10^{-22} ({\rm Z}/{\rm Z_{\odot}}) \mbox{ mag cm}^{2}$ for our choice of dust properties.

To understand these results, first consider only the runs performed
with our default values for $L$ and $G_{0}$. Figure~\ref{mol-AV}a
shows that if we disregard the $L = 5$~pc runs, then a clear 
correlation exists between $\mwfrac{{\rm H_{2}}}$ and $\meanAV$,
with larger values of $\meanAV$ yielding larger values of  
$\mwfrac{{\rm H_{2}}}$. Similarly, if we disregard the $G_{0} = 0.0$
runs, then Figure~\ref{mol-AV}b shows that there is also a clear, and
far stronger,
correlation between $\mwfrac{{\rm CO}}$ and $\meanAV$. 
If we compare the runs
with $\meanAV = 1$  and $\meanAV = 10$, then we see a factor of two 
to three change in $\mwfrac{{\rm H_{2}}}$, compared with
a change in $\mwfrac{{\rm CO}}$ of three to four orders of magnitude.
Plotting $\mwfrac{{\rm H_{2}}}$ and $\mwfrac{{\rm CO}}$ against
the mean gas number density $n_{0}$ shows no clear correlation in
either case (Figure~\ref{mol-n}), if we consider the data as a whole, 
but also shows that both $\mwfrac{{\rm H_{2}}}$ and $\mwfrac{{\rm CO}}$
correlate well with density at fixed metallicity, and with metallicity at
fixed density.

At first glance, these results suggest that both the H$_{2}$ and the CO
abundances are controlled by photodissociation, with the CO 
responding far more strongly to changes in $\meanAV$ than the
H$_{2}$. However, the results of the $L = 5$~pc runs challenge
this interpretation. They lie where we would
expect in the plot of $\mwfrac{{\rm CO}}$ versus $\meanAV$
(Figure~\ref{mol-AV}b), but they lie to the left of the corresponding 
         $L = 20$~pc
runs in the plot of $\mwfrac{{\rm H_{2}}}$ versus $\meanAV$
(Figure~\ref{mol-AV}a). Indeed, shifting them uniformly to the right by a factor
of four brings them into good agreement with the other runs. This result
can be understood if we realize that the real correlation responsible
for the behaviour seen in Figure~\ref{mol-AV}a is not one between 
$\mwfrac{{\rm H_{2}}}$ and  $\meanAV$, but is instead a correlation
between $\mwfrac{{\rm H_{2}}}$ and $n_{0} {\rm Z}$. Since 
$\meanAV \propto n_{0} {\rm Z}$, we find an apparent correlation 
between $\mwfrac{{\rm H_{2}}}$ and  $\meanAV$ if we compare only
runs with the same box size $L$. However, if we change $L$ without
changing $n_{0} {\rm Z}$, we will still obtain the same value of 
$\mwfrac{{\rm H_{2}}}$, but find that it now corresponds to a different
mean visual extinction.

This result strongly suggests that the differences in the mean H$_{2}$ 
abundance found at the end of the different runs are not a result of
photodissociation, and that this plays little or no role in regulating the
amount of H$_{2}$ present in the gas. Instead, the dependence of
$\mwfrac{{\rm H_{2}}}$ on $n_{0} {\rm Z}$ suggests that the amount
of H$_{2}$ present depends simply on the H$_{2}$ formation time,
$t_{\rm form, H_{2}}$, which scales as $t_{\rm form, H_{2}} \propto 1 / (n_{0} {\rm Z})$.
If $n_{0} {\rm Z}$ is small, then the H$_{2}$ formation time is long,
and the hydrogen in the gas simply does not have sufficient time to 
become fully molecular by the end of the simulation, despite the boost 
to the H$_{2}$ formation rate resulting from the presence of the turbulent
compressions. On the other hand, when $n_{0} {\rm Z}$ is large,
the hydrogen can easily become almost fully molecular.

In contrast, the CO abundances in these runs respond in the fashion 
that we would expect them to if they were primarily controlled by 
photodissociation. Decreasing $L$ while holding $n_{0} {\rm Z}$
constant leads to a sharp drop in the CO abundance, implying that
in this case the correlation with $\meanAV$ is real, and not simply
a consequence of the relationship between $n_{0} {\rm Z}$ and
$\meanAV$.

We can confirm this interpretation by considering what happens in the
runs that are performed with $G_{0} = 0.0$ to eliminate all
photodissociation.  We see from 
Figure~\ref{mol-AV}a that in this case, we still recover the same 
correlation between $\mwfrac{{\rm H_{2}}}$ and $\meanAV$. 
However, in the absence of an ultraviolet background, the 
H$_{2}$ abundance {\em cannot} depend on the visual extinction,
as the only physical processes in the simulation that depend on the 
visual extinction are those involving the attenuation of the ultraviolet 
background. Therefore, the correlation between $\mwfrac{{\rm H_{2}}}$ 
and $\meanAV$ cannot be real, and must simply be a consequence
of the actual correlation of between $\mwfrac{{\rm H_{2}}}$ and $n_{0} {\rm Z}$,
or rather the inverse correlation between $\mwfrac{{\rm H_{2}}}$ and
the mean H$_{2}$ formation time.  On the other hand, Figure~\ref{mol-AV}b 
shows that the CO abundance at low $\meanAV$  is much larger in the runs 
without a UV background, consistent with a picture in which the CO abundance 
is primarily  controlled by photodissociation.

Figure~\ref{mol-AV}b also demonstrates that the CO abundance is not 
determined solely by photodissociation, since even in the absence of the
UV background, we see a systematic decrease in the CO abundance as
we move to lower densities. This decrease is driven primarily by the 
decrease in the mean H$_{2}$ abundance, since H$_{2}$ is a key 
component in most of the chemical pathways capable of forming gas-phase
CO. We have verified this by running simulations similar to runs n30-UV0
and n100-UV0 that start with all of the hydrogen in molecular form. The
CO abundances produced in these runs are significantly higher than 
in the runs that start with atomic hydrogen: in the lower density run, roughly
60\% of the available carbon is converted to CO, while in the higher density
run, roughly 95\% is converted. In comparison, the fraction of carbon 
converted to CO in runs n30-UV0 and n100-UV0 is 13\% and 60\%, 
respectively. Nevertheless, it is clear from Figure~\ref{mol-AV}b that the
dependence of the CO abundance on the H$_{2}$ abundance is a relatively 
small effect in comparison with the dependence of the CO abundance on 
the mean visual extinction in the runs with a UV background:  the former
is responsible for changes in $\mwfrac{{\rm CO}}$ of a factor of a few, 
while the latter is responsible for changes of several orders of magnitude.

We have thus arrived at the first key result of this paper: in a 
turbulent molecular cloud, the mean H$_{2}$ abundance depends primarily
on the time taken to form the H$_{2}$ and is insensitive to photodissociation,
while the CO abundance is determined primarily by photodissociation, with
only a secondary dependence on the H$_{2}$ fraction.

\subsection{Molecular abundances: time dependence}
\begin{figure}
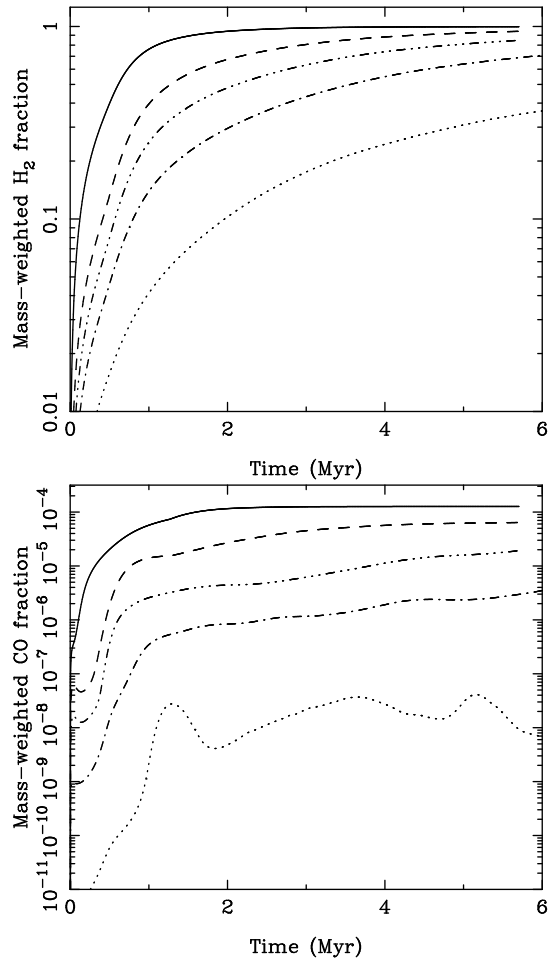

\centering
\epsfig{figure=f4a.eps,width=15pc,angle=270,clip=}
\epsfig{figure=f4b.eps,width=15pc,angle=270,clip=}
\caption{(a) Evolution of the mass-weighted mean H$_{2}$ abundance
with time in runs n1000 (solid line), n300 (dashed line), n180 (dash-dot-dot-dotted line),
n100 (dash-dotted line) and n30 (dotted line).
(b) As (a), but for the mass-weighted mean CO abundance. The large fluctuations
seen in run n30 
         occur because 
the dominant contribution to $\mwfrac{{\rm CO}}$
in this simulation comes from only a few transient CO-rich clumps. As
         the turbulence forms and destroys them,
their contribution to $\mwfrac{{\rm CO}}$ undergoes 
large fluctuations. In the higher density runs, many more structures contribute significantly
to $\mwfrac{{\rm CO}}$ and so the influence of the formation or destruction of any one 
clump is much smaller. \label{mol_time_solar}}
\end{figure}

\begin{figure}
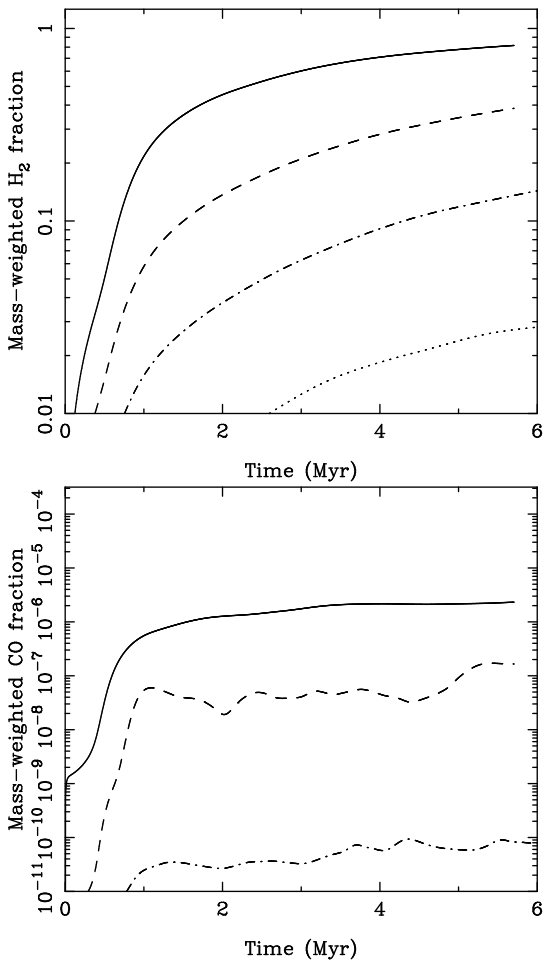

\centering
\epsfig{figure=f5a.eps,width=15pc,angle=270,clip=}
\epsfig{figure=f5b.eps,width=15pc,angle=270,clip=}
\caption{(a) Evolution of the mass-weighted mean H$_{2}$ abundance 
with time in runs n1000-Z01 (solid line), n300-Z01 (dashed line),
n100-Z01 (dash-dotted line)  and n30-Z01 (dotted line). 
(b) Evolution of the mass-weighted mean CO abundance with time in runs 
n1000-Z01 (solid line), n300-Z01 (dashed line), and 
n100-Z01 (dash-dotted line). In run n30-Z01, the mean CO abundance
remains smaller than $10^{-11}$ for the whole period plotted.
Reducing the metallicity clearly has a much more pronounced effect on the
CO than on the H$_{2}$.
\label{mol_time_01}}
\end{figure}

We can get further confirmation of the results of the previous section 
by considering the time dependence of $\mwfrac{{\rm H_{2}}}$ and 
$\mwfrac{{\rm CO}}$. In Figure~\ref{mol_time_solar}, we plot  the evolution 
of these quantities as a function of time for a set of runs with different mean
densities at solar metallicity, while in Figure~\ref{mol_time_01}, we show
a similar plot for runs with a metallicity ${\rm Z} = 0.1 \: {\rm Z_{\odot}}$.
Looking at the evolution of H$_{2}$ fraction with time in these figures,
we see that the time required to convert a large fraction of the initial
atomic hydrogen to molecular hydrogen decreases as we increase 
the density or the metallicity, just as we would expect given the results
of the previous section. Moreover, it is clear that in most of the runs, the
mean H$_{2}$ fraction has yet to reach equilibrium at the end of the run.
Finally, extrapolation of the curves beyond six Myr suggests that the 
H$_{2}$ fraction in all of the runs should eventually reach a value of
order unity; there is no indication that the equilibrium value of 
$\mwfrac{{\rm H_{2}}}$ will be significantly smaller than one in any of 
the runs. 

The time evolution of $\mwfrac{{\rm CO}}$ in the different runs is
qualitatively different from that of  $\mwfrac{{\rm H_{2}}}$. CO forms
rapidly at the beginning of the simulations, but after roughly a million
years, the rate of increase of the CO abundance slows or stops. The
amount of CO formed during this initial period of rapid evolution 
depends strongly on the mean visual extinction of the gas, with the
runs with low $\meanAV$ forming considerably less CO than the
runs with high $\meanAV$. The subsequent evolution of the CO
abundance depends on the value of $\mwfrac{{\rm H_{2}}}$ at 
this point. If $\mwfrac{{\rm H_{2}}} \sim 1$, as in runs n300 or 
n1000, then there is little further evolution of the CO abundance.
On the other hand, if the mean H$_{2}$ abundance remains 
significantly smaller than unity at this early point, then the CO 
abundance continues to increase, driven by the increasing 
availability of H$_{2}$, and may double or triple by the end of 
the run.

It is also evident that $\mwfrac{{\rm CO}}$ displays far larger
fluctuations in runs with low mean visual extinctions (e.g. runs
n30 or n100-Z01) than in runs with high mean visual extinctions.
This is a consequence of the difference in the  spatial distribution 
of the CO in a low extinction run compared to a high extinction run.
When $\meanAV$ is small, the CO fraction in most of the gas is
very small, owing to the lack of effective shielding of the CO against
UV photodissociation. In this case, most of the CO in the simulation
is found in a small number of dense, comparatively well-shielded 
clumps. Since our simulations were performed without self-gravity,
these clumps are transient objects, and are continually being disrupted
and reformed by the turbulence. This continual variation in the number
and nature of the CO-rich clumps leads to the observed variation in
the mass-weighted mean CO abundance. On the other hand, when
$\meanAV$ is large, far more of the gas has a high CO fraction, and
many more clumps and other dense structures such as sheets and
filaments contribute to $\mwfrac{{\rm CO}}$. Although all of these
structures are transient, just as in the low $\meanAV$ runs, the fact
that a  much larger number of structures contribute to $\mwfrac{{\rm CO}}$ 
than in the low $\meanAV$ runs means that the variation in 
$\mwfrac{{\rm CO}}$ resulting from the creation or destruction of
any particular clump or filament is much smaller. 

\begin{figure}
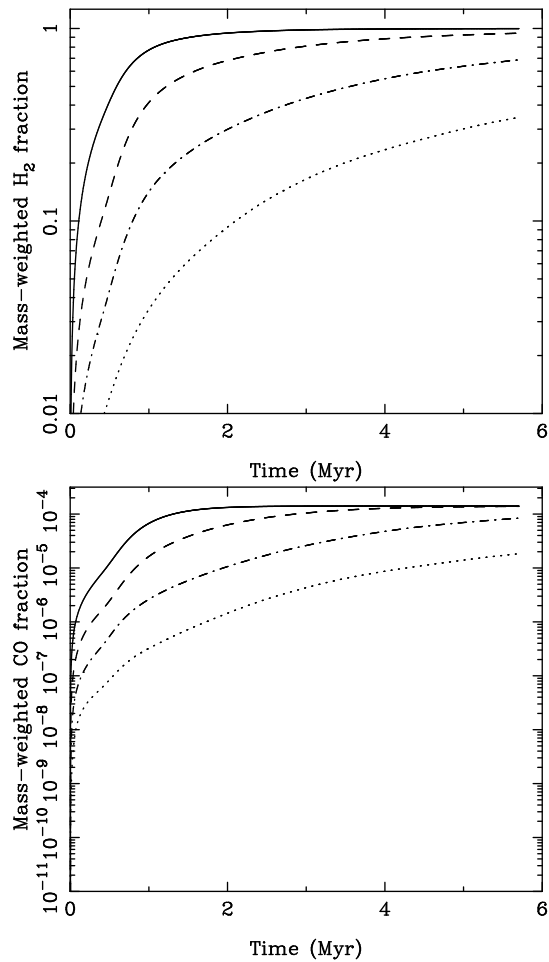

\centering
\epsfig{figure=f6a.eps,width=15pc,angle=270,clip=}
\epsfig{figure=f6b.eps,width=15pc,angle=270,clip=}
\caption{(a) Evolution of the mass-weighted mean H$_{2}$ abundance 
with time in runs n1000-UV0 (solid line), n300-UV0 (dashed line), 
n100-UV0 (dash-dotted line) and n30-UV0 (dotted line).
(b) As (a), but for the mass-weighted mean CO abundance. \label{mol_time_G0}}
\end{figure}

In Figure~\ref{mol_time_G0}, we show how $\mwfrac{{\rm H_{2}}}$ and 
$\mwfrac{{\rm CO}}$ vary with time in runs n30-UV0, n100-UV0, n300-UV0, 
and n1000-UV0, all performed without an ultraviolet background. The
evolution of the H$_{2}$ fraction in these runs is very similar to that in
the runs with a UV background, again showing that photodissociation
plays little or no role in regulating the growth of the H$_{2}$ abundance.
On the other hand, the evolution of the CO abundance in these runs is
clearly different from that in the runs with a UV background. The increase
in the CO abundance in these runs more clearly tracks the increase in the
H$_{2}$ abundance, and there is no longer any significant feature at
$t \sim 1 \: {\rm Myr}$ (other than in run n1000-UV0, which has converted
almost all of its carbon into CO by this point). The CO abundances 
produced in the lower density runs are also clearly much larger than in
the corresponding runs with a UV background, and there is no indication 
that they have yet reached equilibrium.

\begin{figure}
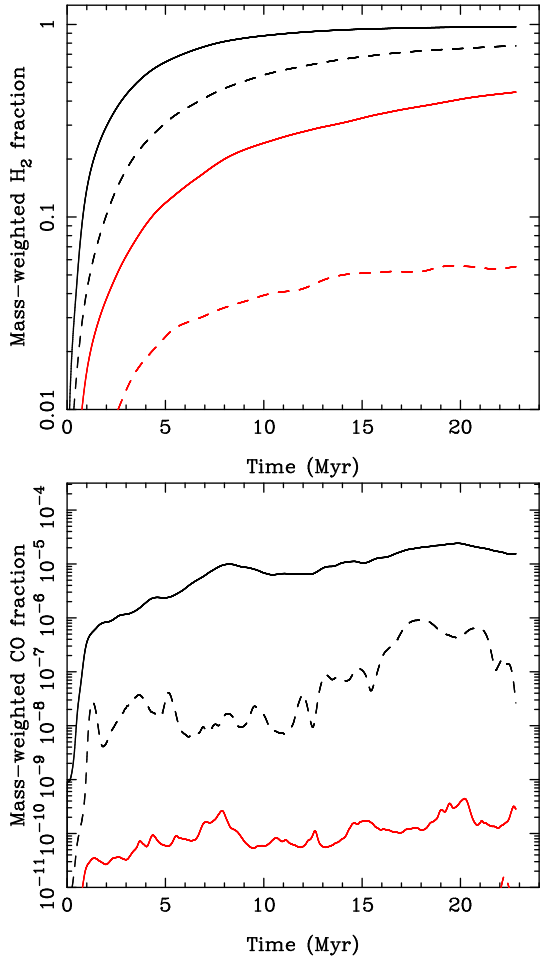

\centering
\epsfig{figure=f7a.eps,width=15pc,angle=270,clip=}
\epsfig{figure=f7b.eps,width=15pc,angle=270,clip=}
\caption{(a) Evolution of the mass-weighted mean H$_{2}$ abundance with time in runs n100
(black solid line), n30 (black dashed line), n100-Z01 (red solid line), and
n30-Z01 (red dashed line).
(b) As (a), but for the mass-weighted mean CO abundance. Note that in run n30-Z01, the
mean CO abundance remains smaller than $10^{-11}$ for almost the whole of the period 
plotted, becoming barely visible in the plot only after $t \sim 22 \: {\rm Myr}$. \label{mol_time_long}}
\end{figure}

Finally, in Figure~\ref{mol_time_long}, we plot the time evolution of 
$\mwfrac{{\rm H_{2}}}$ and $\mwfrac{{\rm CO}}$ in the four runs that we
evolved for a much longer time. This plot further emphasizes that time is the 
primary factor controlling the H$_{2}$ abundance in the runs in which the 
product $n_{0} {\rm Z}$ is small. For instance, at $t \sim 5 \: {\rm Myr}$,
the mass-weighted mean H$_{2}$ fraction in run n100-Z01 is only 10\%,
but it has increased to roughly 40\% by $t \sim 20 \: {\rm Myr}$, i.e.\ it has
increased almost
linearly with time. The only one of these runs in which H$_{2}$ 
photodissociation appears to play an important role is run n30-Z01, 
in which the mean H$_{2}$ fraction appears to saturate at roughly 5\%.
The mean visual extinction of the gas in this run is only 0.1 magnitudes, 
while in run n100-Z01, which is not significantly affected by the UV, it is
approximately 0.3 magnitudes, suggesting that the critical value below
which UV photodissociation begins to significantly limit the H$_{2}$
abundance lies somewhere in the range $\meanAV \sim 0.1$--0.3. This
corresponds to a mean column density of hydrogen nuclei of
$\langle N_{\rm H} \rangle \sim 2$--$6 \times 10^{20} \: {\rm cm^{-2}}$,
which compares well with the finding by the {\em Copernicus} satellite 
that there is a sudden jump in the H$_{2}$ abundance along galactic
sightlines with $\log N_{\rm H} > 20.7$, consistent with the onset of
self-shielding \citep{sav77}. It also agrees within a factor of a few with
the critical extinction $A_{\rm V, crit} \simeq 0.5$ derived by \citet{kmt08}
for idealized spherical clouds.

\subsection{Estimating the CO-to-H$_{2}$ conversion factor}

\begin{figure}
\epsfig{figure=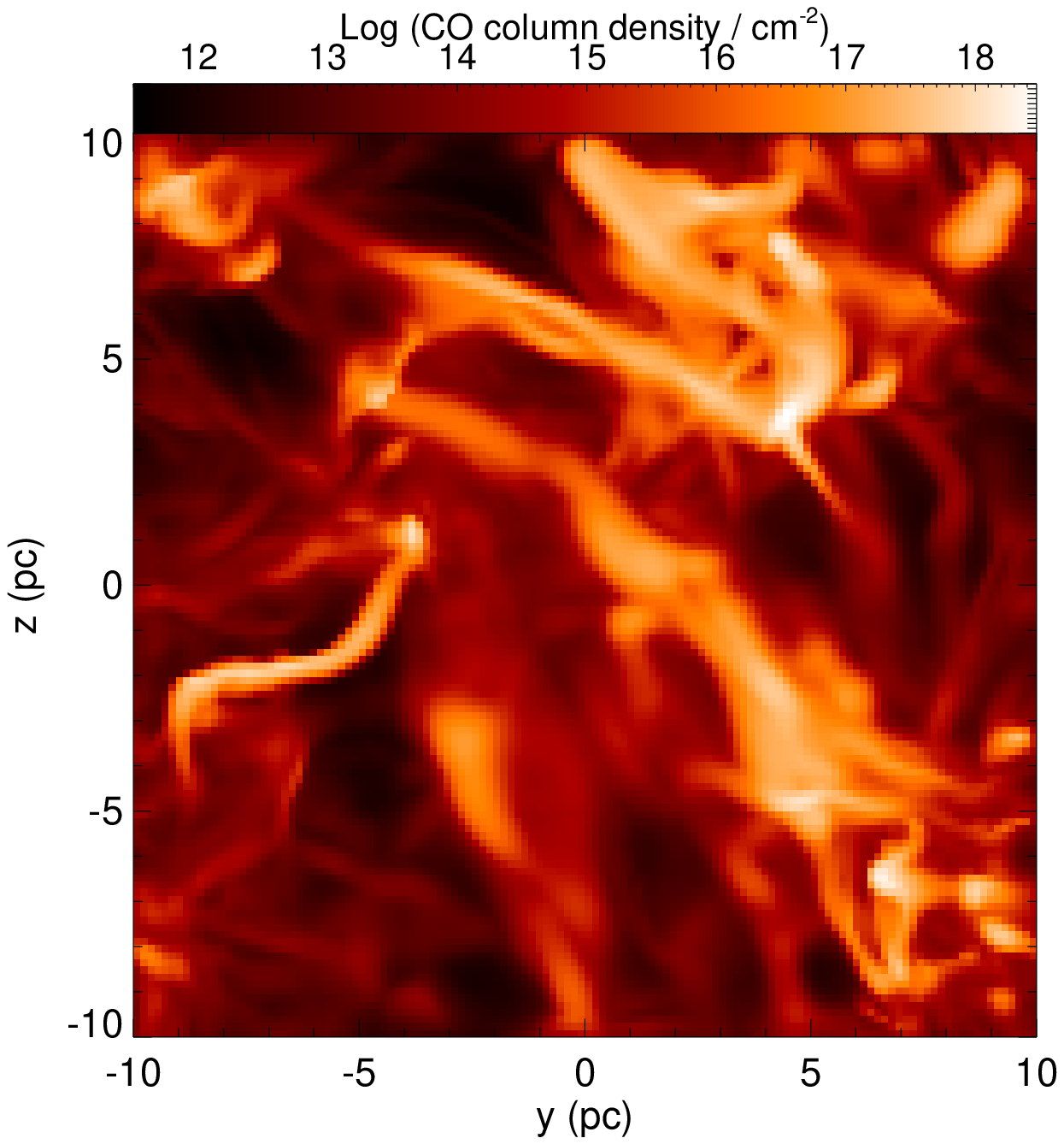,width=20pc,angle=0,clip=}
\epsfig{figure=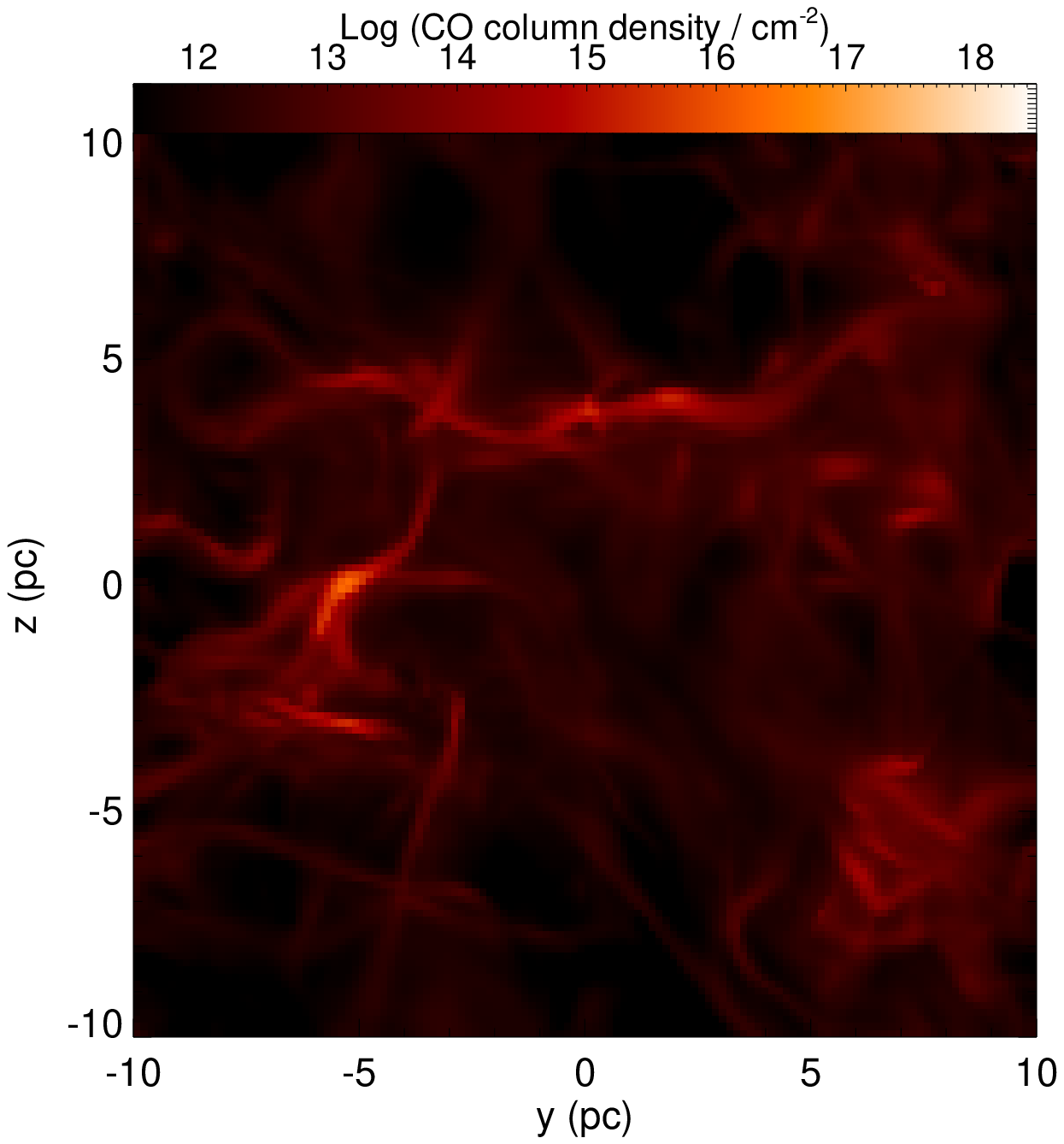,width=20pc,angle=0,clip=}
\caption{(a) CO column density in run n100 at time $t = t_{\rm end}$, viewed along a line
of sight parallel to the $x$-axis of the simulation.
(b) As (a), but for run n30. The contrast with run n100 is striking, and illustrates how
a relatively small change in the mean extinction can produce a large change in the
appearance of the cloud. \label{proj}}
\end{figure}

It is of particular interest to understand how the distinct chemical
changes we see as we decrease $\meanAV$ affect the CO-to-H$_{2}$ conversion factor, $X_{\rm CO}$. 
The decrease of $\mwfrac{{\rm CO}}$ by almost four orders of magnitude
between $\meanAV = 10$ and $\meanAV = 1$, is suggestive of a strong
correlation between $X_{\rm CO}$ and $\meanAV$, but does not by itself
allow us to quantify the relationship between these two
quantities. The sharp decrease in $\mwfrac{{\rm CO}}$ as we decrease
$\meanAV$ is reflected in a sharp decrease in the column density of CO
along most sightlines, as illustrated in Figure~\ref{proj}.  This also
suggests a strong correlation between $X_{\rm CO}$ and $\meanAV$, but
still does not allow us to make a quantitative statement.

To quantify the relationship between $X_{\rm CO}$ and $\meanAV$, we need to be able
to at least approximately compute
\begin{equation}
 X_{\rm CO} \equiv \frac{N_{\rm H_{2}}}{W_{\rm CO}}, 
\end{equation}
where $W_{\rm CO}$ is the velocity-integrated intensity of the $J = 1 \rightarrow 0$
line of $^{12}$CO.  Since we can easily measure $N_{\rm H_{2}}$ along any
desired line of sight through our simulations, the difficulty
comes from the calculation of the integrated intensity. If the gas were optically thin
in the $1 \rightarrow 0$ line, then it would be trivial to compute $W_{\rm CO}$, 
since the CO emissivity at any point would depend only on the local values of the
temperature, density and chemical abundances, and would be effectively decoupled
from the conditions elsewhere in the cloud.

Unfortunately, in many of our simulations, some or all of the lines of sight are
optically thick in the  $1 \rightarrow 0$ line, making it far more challenging to
model the CO emission accurately. When the gas is optically thick, it becomes
necessary to account for line absorption, which can be ignored in the optically
thin case. Furthermore, the opacity and emissivity in the line at
any one point in the cloud can couple to the opacity
and emissivity at other points in the cloud. Therefore, to model the CO emission
entirely accurately, it becomes necessary to calculate the opacities and 
emissivities for all points in the cloud simultaneously. This is typically accomplished
through the use of an iterative method \citep[see e.g.\ the review of techniques
in][]{vz92}. If the geometry of the gas is relatively simple (e.g.\ slab symmetry or spherical 
symmetry), then the resulting problem may still be solved without too much
effort. In the present case, however, the density, temperature,
chemical composition and velocity structure of the gas are all highly inhomogeneous, 
and solution of the coupled equations is a complex and computationally intensive
problem that lies outside of the scope of this initial study. 

If we are prepared to tolerate a certain amount of uncertainty, however, then
it becomes possible to estimate the CO-to-H$_{2}$ conversion factor relatively
easily, even in optically thick gas. To compute an estimate for the conversion
factor, which we denote as $X_{\rm CO, est}$ to distinguish it from the true 
conversion factor $X_{\rm CO}$, we use the following procedure. We begin
by selecting a set of independent lines of sight through our simulation results.
We choose lines of sight that run parallel to the $x$-axis, one per resolution
element.  This choice is arbitrary; a different orientation would yield
similar results. We then compute H$_{2}$ and CO column densities along 
each of these lines of sight. We next convert each of the CO column densities into 
an optical depth, $\tau_{10}$, which represents an estimate of the optical depth of 
the gas in the CO $J = 1 \rightarrow 0$ transition. To perform this conversion, we 
make three major simplifying assumptions. We assume that all of the CO is in 
local thermodynamic equilibrium (LTE), and that the gas is isothermal, with a temperature 
equal to the weighted mean temperature of the gas, where we use the CO number
density as an appropriate weighting function
\begin{equation}
T_{\rm mean} = \frac{\sum_{i,j,k} T(i,j,k) n_{\rm CO}(i,j,k)}{\sum_{i,j,k} n_{\rm CO}(i,j,k)}.
\end{equation}
We also assume
that the CO linewidth $\Delta v$ is uniform, and is given by $\Delta v = 3 \: {\rm km} \:
{\rm s^{-1}}$. Our choice for $\Delta v$ is motivated by the fact that our simulations
have an RMS turbulent velocity of $5 \: {\rm km \: s^{-1}}$, and that we would
expect the one-dimensional velocity dispersion to be a factor of $\sqrt{3}$
smaller than this, i.e.\ $\sigma_{\rm 1D} = 5 / \sqrt{3} \simeq 3 \: {\rm km \: s^{-1}}$.
Given these assumptions, the relationship between $N_{\rm CO}$
and $\tau_{10}$ becomes \citep{tielens05}
\begin{equation}
\tau_{10} = \frac{A_{10} c^{3}}{8 \pi \nu_{10}^{3}} \frac{g_{1}}{g_{0}} f_{0} \left[
1 - \exp\left(\frac{-E_{10}}{kT} \right) \right] \frac{N_{\rm CO}}{\Delta v},
\end{equation}
where $A_{10}$ is the spontaneous radiative transition rate for the 
$J = 1 \rightarrow 0$ transition, $\nu_{10}$ is the frequency of the transition,
$E_{10} = h\nu_{10}$ is the corresponding energy, $g_{0}$ and $g_{1}$ are
the statistical weights of the $J=0$ and $J=1$ levels, respectively, and
$f_{0}$ is the fractional level population of the $J=0$ level. We take
values for $A_{10}$ and $\nu_{10}$ from the LAMDA database \citep{sch05}.
In the case that $T_{\rm mean} = 10 \: {\rm K}$, a typical temperature in the
highly molecular gas in our simulations,  this procedure yields
$\tau_{10} \simeq 5 \times 10^{-17} N_{\rm CO}$, i.e.\  in this case, a CO 
column density of approximately $2 \times 10^{16} \: {\rm cm^{-2}}$ corresponds 
to an optical depth $\tau_{10} = 1$.

To convert from $\tau_{10}$ to $W_{\rm CO}$, we use the same procedure as in
the curve of growth analysis presented in \citet{pineda08}. We write the integrated
intensity as
\begin{equation}
W_{\rm CO} = T_{\rm b} \Delta v \int_{0}^{\tau_{10}} 2 \beta(\tilde{\tau}) {\rm d}\tilde{\tau}, 
\label{wco}
\end{equation}
where $T_{\rm b}$ is the brightness temperature of the line, $\Delta v$ is the
linewidth, and $\beta$ is the photon escape probability. We approximate $\beta$
by assuming that it is the same as for a plane-parallel, uniform slab \citep{tielens05}:
\begin{equation}
\beta(\tau) = \left \{ \begin{array}{lr} [1 - \exp(-2.34\tau)] / 4.68 \tau & \tau \leq 7, \\
\left(4\tau \left[\ln \left(\tau / \sqrt{\pi} \right) \right]^{1/2} \right)^{-1} & \tau > 7.
\end{array} \right.
\label{beta}
\end{equation}
Finally, since we have previously assumed that the CO is in LTE, we simply
set $T_{\rm b} = T_{\rm mean}$.

Using Equations~\ref{wco} and \ref{beta}, we can compute the integrated
intensity for each of our lines of sight. Since we already know the H$_{2}$
column density along each line of sight, we could in principle associate a
CO-to-H$_{2}$ conversion factor with each one. However, as the size  
of each individual zone in one of our simulations is much smaller than the
spatial resolution of observations of all but the few closest GMCs, we feel
it leads to a more meaningful value for $X_{\rm CO, est}$ if we average over 
the whole cloud, since this is essentially what the 
limited resolution of the observations forces us to do observationally.

We therefore average over all lines of sight
to compute a mean intensity $\meanW$ for the datacube as a whole. Similarly,
we compute a mean H$_{2}$ column density for the datacube, $\meanHt$,
by averaging over all lines of sight. Our 
estimate of the CO-to-H$_{2}$ conversion factor is then simply the ratio of these
two quantities:
\begin{equation}
X_{\rm CO, est} = \frac{\meanHt}{\meanW}.
\end{equation}

\begin{figure}
\centering
\epsfig{figure=f9.eps,width=15pc,angle=270,clip=}
\caption{Estimate of the CO-to-H$_{2}$ conversion factor $X_{\rm CO, est}$,
plotted as a function of the mean visual extinction of the gas, $\meanAV$. 
The simplifications made in our modelling mean that each value of 
$X_{\rm CO, est}$ is uncertain by at least a factor of two. At $\meanAV > 3$, 
the values we find are consistent with the value of 
$X_{\rm CO} = 2 \times 10^{20} {\rm cm^{-2} \: K^{-1} \: km^{-1} \: s}$
determined observationally for the Milky Way by \citet{dame01}, 
indicated in the plot by the horizontal dashed line. 
At $\meanAV < 3$, we find evidence for a strong dependence of
$X_{\rm CO, est}$ on $\meanAV$. The empirical fit given by 
Equation~\ref{emp-fit} is indicated as the dotted line in the Figure, and demonstrates that
at low $\meanAV$, the CO-to-H$_{2}$ conversion factor increases roughly
as $X_{\rm CO, est} \propto A_{\rm V}^{-3.5}$. It should also be noted that at 
any particular $\meanAV$, the dependence of $X_{\rm CO, est}$ on
metallicity is relatively small. Previous claims of a strong metallicity
dependence likely reflect the fact that there is a strong dependence
on the mean extinction, which varies as $\meanAV \propto {\rm Z}$
given fixed mean cloud density and cloud size. 
\label{xfact}
}
\end{figure}

\begin{figure}
\centering
\epsfig{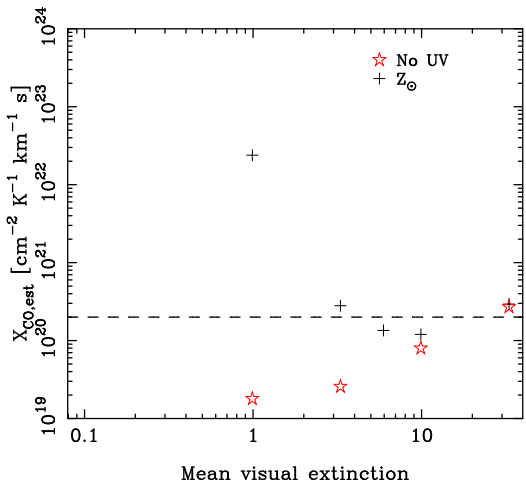}
\caption{Estimate of the CO-to-H$_{2}$ conversion factor $X_{\rm CO, est}$,
plotted as a function of the mean visual extinction of the gas, $\meanAV$. In
this Figure, we compare the values obtained for our solar metallicity runs
with and without the presence of an ultraviolet background. At high 
$\meanAV$, the presence of a UV background has little effect on our
derived value of $X_{\rm CO, est}$, but at low $\meanAV$, CO photodissociation
plays an important role in producing the observed Milky Way value of
$X_{\rm CO} = 2 \times 10^{20} {\rm cm^{-2} \: K^{-1} \: km^{-1} \: s}$
\citep{dame01}, indicated here by the horizontal dashed line.
\label{xfact-G0}
}
\end{figure}

Using this procedure, we have computed this estimated CO-to-H$_{2}$ conversion factor 
for each of  our simulations, and have plotted the resulting values against $\meanAV$ in
Figure~\ref{xfact}--\ref{xfact-G0}. In Figure~\ref{xfact}, we show the results from all of our
runs with non-zero UV backgrounds. In Figure~\ref{xfact-G0}, we compare our results from 
the runs no UV with those from the corresponding runs with UV.
The canonical value of $X_{\rm CO}$ for the Milky Way, 
$X_{\rm CO} = 2 \times 10^{20} {\rm cm^{-2} \: K^{-1} \: km^{-1} \: s}$ 
\citep{dame01} is indicated in both Figures by
a horizontal dashed line. 

The first point to note from Figure~\ref{xfact} is that for
our solar metallicity runs with $\meanAV > 3$, or in other words, for
simulations with parameters that are a reasonable match to the properties
of local GMCs, we find remarkably good agreement between our estimated
conversion factor and the measured value.  This level of agreement is probably
fortuitous to some degree, given the uncertainty in the observational value and
the larger uncertainties in our estimated values. Nevertheless, it does indicate that
despite our simplifications, our estimates do seem to be successfully capturing the
basic physics and are probably trustworthy to within a factor of a few.

A second important point to note from Figure~\ref{xfact} is that for any given 
$\meanAV$, the CO-to-H$_{2}$ conversion factor shows little dependence
on metallicity -- changes to ${\rm Z}$ of an order of magnitude or more alter
$X_{\rm CO, est}$ by at most a factor of a few. It is therefore not true to say
that $X_{\rm CO}$ has a strong dependence on metallicity. Instead, Figure~\ref{xfact}
demonstrates that the true dependence is on the mean extinction
$\meanAV$. For $\meanAV \simgreat 3$, there is no clear correlation between
$X_{\rm CO, est}$ and $\meanAV$, but at $\meanAV \simless 3$, we see a strong
correlation. The dependence of  $X_{\rm CO, est}$ on $\meanAV$ can be described
with the following, wholly empirical fitting function, indicated in Figure~\ref{xfact} by
the dotted line:
\begin{equation}
X_{\rm CO, est} \simeq \left \{
\begin{array}{lr}
2.0 \times 10^{20} & A_{\rm V} > 3.5 \\
2.0 \times 10^{20} \left(A_{\rm V} / 3.5\right)^{-3.5} 
& A_{\rm V} < 3.5
\end{array}
\right .
\label{emp-fit}
\end{equation}
where $X_{\rm CO, est}$ has its standard units of ${\rm cm^{-2}} \: {\rm K^{-1}}
\: {\rm km^{-1}} \: {\rm s}$. This empirical fitting function is accurate to within a factor of a few
over the whole range of  $\meanAV$ and $X_{\rm CO, est}$ considered in
the Figure.

Of course, if we consider clouds with a fixed density and size, and merely
vary the metallicity, then the strong dependence of $X_{\rm CO, est}$ on
the mean extinction implies a strong metallicity dependence, since 
$\meanAV$ is proportional to metallicity. On the other hand, if we vary
the metallicity but also allow the mean density or the size of the cloud to
vary to compensate, then we will find little or no metallicity dependence.

Figure~\ref{xfact-G0} demonstrates the importance of UV photodissociation
for producing the Galactic CO-to-H$_{2}$ conversion factor. In the absence
of UV photodissociation, we get comparable values of $X_{\rm CO, est}$
for our $\meanAV \simeq 10$ and $\meanAV \simeq 30$ runs, as one would
expect, given that most of the gas in these runs is well shielded from whatever
UV is present, but we find significantly smaller values for $X_{\rm CO, est}$ 
at lower mean extinctions. 

\begin{figure}
\centering
\epsfig{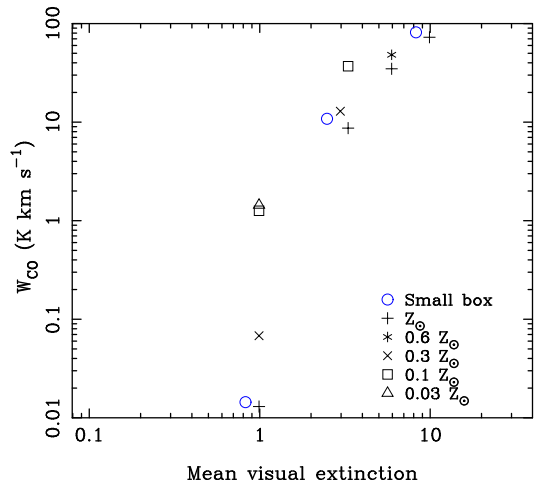}
\caption{Velocity-integrated intensity of the $J=1 \rightarrow 0$ transition
in $^{12}$CO, plotted as a function of the mean extinction of the gas. The
symbols are the same as in our previous plots. Noise levels for extragalactic
CO observations are typically of the order of $1 \: {\rm K} \: {\rm km} \: {\rm s^{-1}}$,
so only the runs in the top right-hand quadrant of the plot produce enough CO
to be readily detectable. 
\label{mean-WCO}
}
\end{figure}

It is also of interest to plot the mean values of $W_{\rm CO}$ for the
various runs, as this gives a good indication of which of our
simulations would correspond to CO-bright clouds, from which it would
be easy to detect CO emission, and which would produce CO-faint clouds
that would be difficult or impossible to detect in CO. In
Figure~\ref{mean-WCO}, we therefore plot $\langle W_{\rm CO} \rangle$
against $\meanAV$ for each of our simulations. Given that the noise
level for observations of CO in Local Group galaxies is typically of
the order of $1 \: {\rm K} \: {\rm km} \: {\rm s^{-1}}$
\citep[e.g.][]{ros07}, it is clear that only the runs with $\meanAV
\simgreat 1$ will produce CO-bright clouds; runs with lower $\meanAV$
will produce clouds that would be extremely difficult to detect with
CO observations. Moreover, of the runs with $\meanAV \sim 1$, only runs
n1000-Z003 and n300-Z01 produce detectable CO: the higher metallicity, lower 
density runs with $\meanAV \sim 1$ yield CO integrated intensities that would be
undetectable.

Finally, it is interesting to speculate how our results for
$X_{\rm CO}$ would be affected if we were to consider stronger UV
field strengths. In clouds with a high mean extinction, we
would not expect the increased UV to have much effect. In
these runs, most of the CO is situated in gas that is already
well shielded from the UV background, and even relatively
large changes in the strength of the background will not lead
to CO photodissociation becoming important in these regions.
On the other hand, in runs with low mean extinction, that have
most of their CO in gas that is not particularly well shielded
from the UV background, a change in the strength of the
background will have a much stronger effect. We would therefore
expect the values we obtain for $X_{\rm CO}$ in the $\meanAV \simgreat 3$ 
runs to have little sensitivity to the UV background field strength, and the
values we obtain in the $\meanAV \simless 3$ runs to increase significantly
with increasing UV field strength, leading to a steepening of
the relationship between $X_{\rm CO}$ and $\meanAV$. We plan to test this
prediction in future work.

\section{Discussion}
\label{dis}
We can draw a number of immediate conclusions from the results that we have
presented in the previous section. First, it seems plain that when we talk
about the formation of a molecular cloud, we should draw a distinction between
the formation of a cloud in which the hydrogen is primarily in molecular form 
and the formation of a cloud from which CO emission can readily be
detected,
since the conditions for the former are not the same as those for the latter.

To form an H$_{2}$-dominated cloud, the cloud must have a mean
visual extinction of a few tenths of a magnitude (cf.\ \citealt{kmt08}, who
find a critical value of approximately 0.5 for a spherical molecular cloud).
More importantly, however, the cloud must survive for long enough
to convert most of its atomic hydrogen into H$_{2}$. The time required is a
strong function of $n_{0} {\rm Z}$, the product of the mean density of the
cloud and its metallicity, although the dependence is not linear, and the
time required also depends upon the strength and nature of the turbulence
(\citealt{gm07b}; Milosavljevic et al., in prep.)

On the other hand, to make a CO-bright cloud, we need a significantly 
higher mean extinction, $\meanAV \simgreat 3$, but do not require so 
much time, since the CO abundance in most of the gas comes 
into equilibrium within 1--2~Myr. In practice, given plausible cloud sizes, clouds
that have a high enough mean extinction to be CO-bright will also be
dense enough to form H$_{2}$ relatively quickly, but the converse is not
necessarily true.

Second, the existence of a visual extinction threshold above which
clouds become CO-bright provides a simple explanation for the
observation that in low metallicity systems cloud masses derived from CO
observations are significantly smaller than those derived from
techniques that do not depend on CO, such as infrared emission
\citep[e.g.][]{israel97, leroy07, leroy09}. In these systems, the CO 
largely traces the regions of a cloud (or cloud complex) that have 
mean extinctions greater than 2, but does not trace the H$_{2}$ in the 
surrounding envelope, which may be considerably more extensive and 
may contain a large fraction of the total mass of the cloud. 

Third, the fact that the CO fraction typically reaches equilibrium on
a timescale comparable to or shorter than the crossing time of the
cloud, and significantly shorter than the time required to assemble
the cloud from warm atomic gas implies that CO emission will
``switch-on'' rapidly during the assembly of the cloud, as
hypothesized by \citet{hbb01}. In the time it takes compressive flows
or gravity to double or triple the mean extinction of a cloud with
$\meanAV \sim 1$ its CO content and luminosity can increase more than
a hundredfold. The cloud will therefore quickly move from being
effectively unobservable in CO to being readily observable.

On the other hand, the long H$_{2}$ formation timescale implies that
molecular clouds need not have equilibrium H$_{2}$ abundances, 
even once they become CO-bright, contrary to what is often assumed
\citep[see e.g.][]{kmt08,mk10}. How far from equilibrium the H$_{2}$
fractions are will depend on the cloud properties (mean density,
metallicity, rms turbulent velocity, etc.), the age of the cloud, and 
also on the range of densities considered within the cloud, since 
overdense regions will reach equilibrium much faster than underdense
regions.

Fourth, we would expect the molecular clouds (or regions
thereof) in low metallicity systems that we observe to be
CO-bright to also be systematically larger and/or denser than their 
counterparts in the Milky Way. This conclusion follows from the fact
that if $\meanAV \simgreat 3$ is required for CO to form, and we decrease ${\rm Z}$,
then we must increase either $n_{0}$ or $L$ to compensate, since
$\meanAV \propto n_{0} L {\rm Z}$. This conclusion also lends itself
to a relatively simple observational test. If we observe a giant molecular
cloud of size $L$ in a low metallicity system (where $L$ is the total
size of the cloud, not of the CO-bright region), and compare it with
a cloud of a similar size in the Milky Way, then the former should be
systematically denser than the latter. This systematic difference in
density may be detectable by examination of the CO(2-1)/CO(1-0)
or  CO(3-2)/CO(1-0) line ratios, or the ratio of of HCN emission to CO 
emission, all of which will be larger in denser systems.

Finally, our results also suggest a relatively simple explanation
for the fact that GMCs observed in the Milky Way span only a
small range in column densities \citep{blitz07}. The requirement
that $\meanAV \simgreat 3$ for a CO-bright cloud implies that 
any cloud identified as a GMC must have a minimum
mean column density of around 
$60 \mbox{ M}_{\odot} \mbox{ pc}^{-2}$,
since clouds with lower column densities will typically not contain
much CO. 

The distribution of the mean column densities of GMCs above this 
minimum value depends upon the distributions of the size and mean 
density of the clouds. However, the cloud size distribution is typically found to
be a sharply decreasing function of cloud size $L$.  For example,
\citet{hcs01} find $n(L){\rm d}L \propto L^{-3.2 \pm 0.1} {\rm d}L$ for GMCs
in the outer Galaxy. If the mean density of the GMCs spans only a
small range of values, then the distribution of column densities
$n(\Sigma){\rm d}\Sigma$ will be a sharply decreasing function of
$\Sigma$. If we combine this fact with the existence of a minimum
column density threshold for a CO-bright GMC, then it implies
that most observed GMCs will have column densities that are
close to the threshold.

This argument relies
on the mean cloud density not being a strong function of the size
of the cloud. It is possible to show that this is a good assumption
for at least the \citet{hcs01} cloud sample. Given a cloud size
distribution $n(L) \propto L^{-3.2 \pm 0.1}$, and a constant
cloud mean density, then the cloud mass distribution is simply
\begin{eqnarray}
n(M){\rm d}M  & = & n(L) \frac{{\rm d}L}{{\rm d}M} {\rm d}M, \nonumber  \\
 & \propto & L^{-3.2 \pm 0.1} L^{-2}  \nonumber \\
 & \propto & M^{-(5.2 \pm 0.1) / 3} \nonumber \\
 & \propto & M^{-1.73 \pm 0.03},
\end{eqnarray}
where we have assumed only that $M \propto L^{3}$, consistent
with our adoption of a constant cloud mean density. For comparison,
the cloud mass distribution inferred by \citet{hcs01} for this sample
of clouds was $n(M) \propto M^{-1.8 \pm 0.03}$, demonstrating that
even if the mean density of the clouds is not constant, it can at most
be a very weak function of cloud size $L$.

If this argument is correct, it suggests that most Galactic GMCs should
have column densities close to 
$60 \: {\rm M_{\odot}} \: {\rm pc^{-2}}$.
This value is a factor of three smaller than the value of 
$170 \: {\rm M_{\odot}} \: {\rm pc^{-2}}$ determined by
\citet{sol87} that is often taken to be canonical, but is in good agreement
with the value recently derived by \citet{hkdj09} from their re-examination 
of the \citeauthor{sol87} clouds using  new data from the Boston University-FCRAO 
Galactic Ring Survey.  

We note in conclusion that this argument makes no assumptions about
the dynamical equilibrium of the clouds.  This casts doubt on the
frequently-made assumption that the small scatter in observed GMC
column densities implies that the GMCs are in virial equilibrium.

\section{Caveats}
\label{cave}
There are several caveats that should be borne in mind when considering our
results. In our simulations, we use a simplified chemical network for modelling
the formation of H$_{2}$ and CO, and adopt an approximate
treatment of the effects of the ultraviolet background radiation field. 
When modelling the CO integrated intensity, $W_{\rm CO}$, we make additional 
approximations, such as the assumption that all of the CO is in LTE, or that the CO 
linewidth has a fixed value. Finally, our simulations neglect the effects of self-gravity, 
which is clearly important in real molecular clouds.

However, we are confident that these issues will not significantly affect our
major results. In Paper I, we showed that our simplified chemical model produces
results that 
         agree excellently 
with the standard UMIST astrochemical
model \citep{umist99} over a wide range of parameter space, provided that H, 
He, C and O are the only elements considered. Including other elements in the
chemical model, such as sulphur, will worsen the agreement slightly, particularly
in regions dominated by C and C$^{+}$, but is unlikely to result in large changes
in either the mean abundance of CO or the CO integrated intensity.

Our simplified treatment of the effects of UV photodissociation will also introduce
some error into our predicted CO abundances. Because the dense gas in a 
molecular cloud has a relatively small volume filling factor, we can be reasonably
confident that our six-ray treatment of the radiation field will tend to overestimate 
its strength, as it will be more likely to miss shielding coming from dense regions 
that do not lie on lines of sight parallel to the coordinate axes, than to overestimate
the effectiveness of the shielding provided by dense regions that do lie along these
lines of sight. However, the main contribution to the photodissociation rate in any 
particular zone will come from the lines of sight with the lowest column densities. 
Therefore, if the lines of sight that run parallel to the coordinate axes have low column 
densities, the fact that we miss the shielding from a few dense regions that do not lie 
along these lines of sight will not substantially alter the predicted photodissociation 
rates.

The approximations that we have made in order to derive $W_{\rm CO}$ are
another potential source of error. Observationally, we know that not all of the
CO in molecular clouds is in LTE \citep{goldsmith08}, and so our assumption 
that it is will cause us to overestimate the integrated intensity. Our assumption
that all of the CO has the same temperature is another source of error, although
we have attempted to mitigate this by adopting a suitably-weighted mean 
temperature for the gas that gives rise to the CO emission.
Our adoption of a fixed linewidth for the CO is also unrealistic:
in reality, CO in lower metallicity or lower density systems will 
         be
distributed more 
intermittently, and so can be expected to have a lower velocity dispersion
than the gas as a whole. However, none of these sources of error will result in
a large change in $W_{\rm CO}$, and moreover they will offset each other to
some degree.

Finally, our neglect of the effects of self-gravity is important if one is concerned
with the structure of the cloud on the scales of individual dense clumps 
\citep[e.g.][]{klessen01,kainulainen09}. However, the quantities that we have 
been dealing with in this paper are the results of averaging over the whole cloud, 
and changes in the cloud structure on scales of a few tenths of a parsec are 
unlikely to significantly affect these cloud-averaged values. Nevertheless, this 
is an issue that we intend to return to in the future.

\section{Summary}
\label{summ}
We finish our discussion here by summarizing the key results of this paper.
We have shown that the H$_{2}$ abundance in turbulent molecular clouds
is controlled primarily by the time taken to form the H$_{2}$, and that it is
relatively insensitive to the effects of UV photodissociation. Photodissociation
becomes significant for determining the H$_{2}$ abundance only below a
visual extinction threshold of a few tenths of a magnitude, in good agreement
with previous observational and theoretical work. On the other hand, CO forms
rapidly, but is strongly affected by photodissociation. The mean CO abundance 
falls off rapidly with decreasing mean extinction, particularly in clouds with 
$\meanAV \simless 3$.

We have demonstrated that the CO-to-H$_{2}$ conversion factor is also 
determined primarily by the mean extinction of the cloud, and that it is almost
constant for $\meanAV \simgreat 3$, but falls off as $X_{\rm CO} \propto
A_{\rm V}^{-3.5}$ for $\meanAV \simless 3$. Furthermore, only the clouds with 
visual extinctions greater than a few have sufficiently large CO integrated 
intensities to be detected by current observations. This therefore suggests
a simple explanation for the discrepancy observed in low metallicity systems
between cloud masses determined by CO observations and those determined 
by non-CO tracers, such as IR emission.  CO observations inevitably select
CO-bright clouds that have mean extinctions large enough to place them in
the regime where $X_{\rm CO}$ is approximately constant and equal to the
Galactic value. On the other hand, observations that do not rely on CO have 
no such bias and so select clouds that lie on the power-law portion of the 
relationship. Therefore, CO observations find an $X_{\rm CO}$ that does
not vary significantly with metallicity, while other observations find a strong
dependence on metallicity.

Finally, we have shown that if we combine the requirement that
$\meanAV \simgreat 3$ for a CO-bright cloud with the observational
fact that the GMC size distribution is a steeply decreasing function of
size, then we are lead quite naturally to the prediction that all GMCs
should have near-constant column densities. Therefore, contrary to what 
is often assumed, the small scatter in observed GMC column densities does 
not necessarily imply that the GMCs are in virial equilibrium.

\section*{Acknowledgements}
The authors would like to thank A.~Bolatto, P.~Clark, L.~Hartmann,
A.~Hughes, F.~Israel, A.~Leroy, M.~Peeples, A.~Sternberg, and W.~Wall
for their comments on 
the work presented in this paper. They also thank the anonymous referee
for a number of suggestions that have helped to improve the paper.
SCOG thanks the 
Deutsche Forschungsgemeinschaft 
(DFG) for its support via grants KL1358/4 and KL1358/5. In addition, he 
acknowledges partial support from a Frontier grant of Heidelberg University,
funded by the German Excellence Initiative, and from the German
Bundesministerium f\"ur Bildung und Forschung 
via the
ASTRONET project STAR FORMAT (grant  05A09VHA). M-MML 
         thanks the Max-Planck-Gesellschaft and the Deutsche
         Akademische Austausch Dienst for support to visit Heidelberg,
         and the National Science Foundation for its support via grant AST 08-35734.

\end{document}